\documentclass[fleqn,usenatbib]{mnras}

\usepackage{newtxtext,newtxmath}

\usepackage[T1]{fontenc}

\DeclareRobustCommand{\VAN}[3]{#2}
\let\VANthebibliography\thebibliography
\def\thebibliography{\DeclareRobustCommand{\VAN}[3]{##3}\VANthebibliography}

\usepackage{graphicx}	
\usepackage{amsmath}	
\usepackage{multirow}

\title[Astrometric Approach to measuring color]{An Astrometric Approach to Measuring the Color of an Object}

\author[Guo et al.]{
	B. F. Guo$^{1,3}$,
	Q. Y. Peng$^{1,3}$\thanks{E-mail: tpengqy@jnu.edu.cn},
	X. Q. Fang$^{1,3}$,
	and F. R. Lin$^{2,3}$
\\
$^{1}$Department of Computer Science, Jinan University, Guangzhou 510632, China\\
$^{2}$School of Software, Jiangxi Normal University, Nanchang 330022, China\\
$^{3}$Sino-French Joint Laboratory for Astrometry, Dynamics and Space Science, Jinan University, Guangzhou 510632, China}

\date{Accepted 2023 Aug 28; Revised 2023 Aug 05; in original form 2023 Jun 09}

\pubyear{2023}

\begin{document}
\label{firstpage}
\pagerange{\pageref{firstpage}--\pageref{lastpage}}
\maketitle

\begin{abstract}
The color of a star is a critical feature to reflect its physical property such as the temperature. The color index is usually obtained via absolute photometry, which is demanding for weather conditions and instruments. In this work, we present an astrometric method to measure the catalog-matched color index of an object based on the effect of differential color refraction (DCR). Specifically, we can observe an object using only one filter or alternately using two different filters. Through the difference of the DCR effect compared with reference stars, the catalog-matched color index of an object can be conveniently derived. Hence, we can perform DCR calibration and obtain its accurate and precise positions even if observed with \textit{Null} filter during a large range of zenith distances, by which the limiting magnitude and observational efficiency of the telescope can be significantly improved. This method takes advantage of the DCR effect and builds a link between astrometry and photometry. In practice, we measure the color indices and positions of Himalia (the sixth satellite of Jupiter) using 857 CCD frames over 8 nights by two telescopes. Totally, the mean color index $ BP-RP $ (Gaia photometric system) of Himalia is 0.750 $\pm$ 0.004 magnitude. Through the rotational phased color index analysis, we find two places with their color indices exceeding the mean $\pm$ 3$\sigma$.
\end{abstract}


\section{Introduction}
\label{section1}

\par The color of a star, which is usually described as the value of color index, contains the key information of its basic property. For example, the color of a star reflects its temperature, and helps us further derive its mass, brightness and metallicity (we only refer to main sequence stars here). The color of a solar system object helps us reveal its composition. It is expected that the color properties of Trans-Neptunian objects (TNOs) and irregular satellites are similar, since these satellites are expected to be the captured TNOs during the early history of the giant planet dynamical evolution \citep{Graykowski2018AJ,Jose2022AJ}. In addition, investigating the color variations of the objects in solar system helps us analyze their atmospheric changes and surface material distributions. For ground-based astrometry, the effect of differential color refraction (DCR) can be calibrated effectively via the color index of an object.

\par The atmospheric refraction occurs when the lights of stars pass through the Earth's atmosphere. In a taken CCD frame by a ground-based telescope, except for the effect of atmospheric refraction, different positional refraction still exists in the star images due to their different colors, which is called DCR. The DCR effect makes the star images in a frame with different positional deviations towards zenith, which increases with the angle away from zenith. The light with shorter wavelength (blue stars) is more refracted, whereas longer wavelength (red stars) less. For ground-based astrometry, the DCR effect results in a systematic positional error, which should be considered to obtain accurate and precise positions \citep{Anderson2006AA, Velasco2016MNRAS}.

\par At present, several methods have been proposed to calibrate the DCR effect in astrometry. For example, \citet{Monet1992AJ} obtained a series of observations when the targets are located at different zenith distances (ZDs, which equal 90$ ^\circ $ - altitude) to determine the DCR effect. Then, they performed absolute photometry to obtain the colors of stars so that the DCR effect can be calibrated. \citet{Velasco2016MNRAS} and \citet{Ducourant2008AA} are the good examples of calibrating the DCR effect using this method. \citet{Stone2002PASP} used H$\alpha$ interference filter (with very narrow passband) to obtain observations, which can be regarded as non-DCR-affected. He also obtained the observations of the same fields of view (FOV) using other filters. Thus, the DCR effect can be calculated when they were mapped onto the non-DCR-affected observations. In this way, the DCR effect can be calibrated after obtaining the colors of stars. Recently, the catalog-based methods have been proposed and one can calibrate the DCR effect based on the astrometric and photometric information provided by some star catalog. For example, \citet{Magnier2020ApJS} calibrated the DCR effect of stars from Pan-STARRS catalog based on the astrometric information from 2MASS catalog \citep{Skrutskie2006AJ}. And later, the star positions of Pan-STARRS catalog have been further improved after the DCR calibration using the astrometric information of Gaia EDR3 catalog \citep{GaiaEDR3_2021AA}. \citet{Lin2020MNRAS} proposed a convenient method using both astrometric and photometric information from Gaia DR2 catalog \citep{GaiaDR2_2018AA}.

\par However, the methods mentioned above mainly focus on the DCR calibration of stars. As for the objects in solar system whose information is not provided by star catalogs, the DCR effect is usually considered in the following ways. Some scholars, for example, \citet{Kilic2012MNRAS}, \cite{Dieterich2018ApJ} and \cite{Shang2022AJ} used the filter with longer wavelength passband to reduce the DCR effect. In addition, the DCR effect can be minimized by restricting the target being close to the meridian when observed. \citet{Ortiz2017Natur}, \citet{Winters2017AJ} and \citet{White2022AJ} reduced the astrometric errors in this way. However, even if we observe close to the meridian, some systematic errors exist in declination. Sometimes, because the astrometric accuracy of moving targets was limited, some works \citep{Gomes2015AA, Wang2015MNRAS,Yu2018PSS} didn't take the DCR effect into account. Of course, there remain some works that calibrated this effect. For example, \citet{Benedetti-Rossi2014AA} observed Pluto using several different filters to calibrate the effect of both reference stars and the target.

\par To accurately calibrate the DCR effect for the solar system objects using the catalog-based method, their catalog-matched color indices should be acquired, however, which are not provided by a star catalog. To obtain the color index of such a target, we can consult some references and transform them to the catalog-matched photometric system. For example, we need to transform to the color index \textit{BP-RP} in Gaia photometric system. For more details of the color index \textit{BP-RP}, one can refer to \citet{Carrasco2021AA}. This transformation may cause large errors (e.g. 0.02 $\sim$ 0.20 mag according to different photometric systems\footnote{https://gea.esac.esa.int/archive/documentation/GDR3}). However, some colors of the targets may have not been published, such as the newly-discovered objects in solar system. Another approach  is to perform absolute photometry, which is demanding for weather conditions (in photometric night) and well-calibrated instruments.

\par In this work, we want to derive the color of a solar system object based on the DCR effect. Since the color index of an object can be described as the function of its DCR effect, we can accordingly derive its color if its DCR effect has been determined. However, it is not easy to acquire the pure DCR effect, because the DCR effect exists when observing with any filter for ground-based telescopes. We will explore this method to obtain the catalog-matched color index of an object in an astrometric way. Once the color index has been derived, we can observe with \textit{Null} or \textit{Clear} filters during a large range of ZDs and perform DCR calibration. Compared with using the filters with long wavelength passband which cause negligible DCR effect \citep{Stone2002PASP, Lin2020MNRAS}, the limiting magnitude and the observational efficiency can be well improved. Compared with the observational scheme of only observing near the meridian, we have much more time to observe a program object. On the other hand, we don't need to perform absolute photometry to acquire the color of an object, which relies on the demanding weather conditions, instruments and photometric system transformation.

\par The contents of this paper are arranged as follows. Section \ref{section:2} will elaborate the method to measure the color of an object using DCR effect. Section \ref{section:3} will introduce the observations used. Section \ref{section:4} will show the measurement results in detail. Section \ref{section:5} will show the discussions, and the conclusions will be drawn in the last section.

\section{Method}
\label{section:2}

\par There are two observation schemes to measure the catalog-matched color index in an astrometric way. One is observing an object only with one filter, and another is alternately using two different filters. The details of the two schemes will be elaborated in this section.

\begin{figure*}
	\centering
	\includegraphics[width=0.97\textwidth]{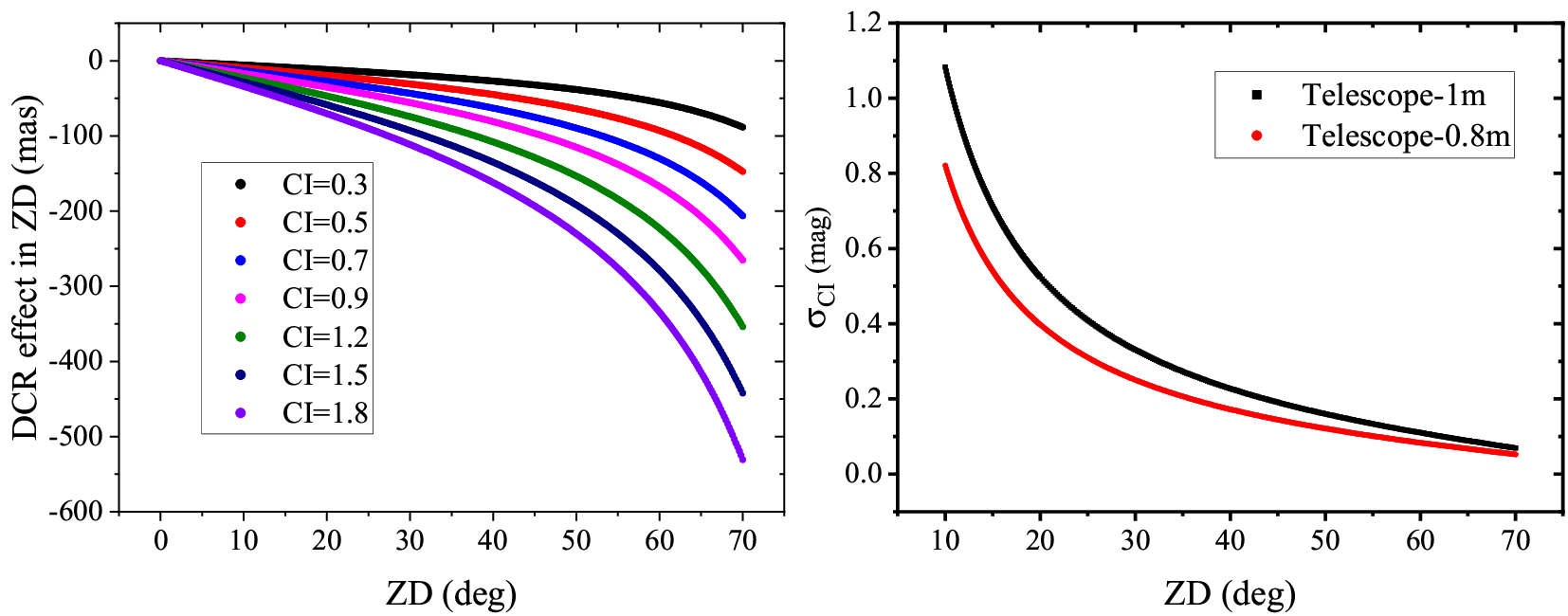}
	\caption{The left panel shows the DCR effect of an object with different color indices \textit{BP-RP} in ZD according to the mean DCR coefficient of \textit{Null} filter over 4 nights by the 1 m telescope. We suppose the DCR effect is zero at zenith. The curves in different colors show the object with different color indices. The right panel shows the measured uncertainty of color index with ZD for an individual frame using two filters alternately. The estimation is according to the mean DCR coefficients and estimated positional uncertainty of Himalia in this work by the two telescopes used.}
	\label{fig:CI_errors}
\end{figure*}

\subsection{Atmospheric Refraction and DCR Effect}
\label{section:2.1}
\par The lights of stars through the Earth's atmosphere cause the positional refraction, including atmospheric refraction and DCR effect. To clearly describe the positional effects of Earth's atmosphere here, the positional effects are divided into color-related effect (called the DCR effect) and color-unrelated effect (called atmospheric refraction), although the atmospheric refraction includes the differential color refraction effect in some references. Here, we refer to The Astronomical Almanac \citep{Bell2012Book}, according which the atmospheric refraction is not relevant to light wavelength. As for DCR effect, the lights of stars will refract with different positions towards zenith according to their different light wavelengths. For ground-based telescopes, the DCR effect exists when observing with any filter (even more  obvious effect without using filter). For an observed star image, the observational position in the direction of ZD can be expressed via the following equation (this equation is applicable for any filter in optical band or no filter in optical telescope),
\begin{equation}
	\Delta z_{(O-C)} = \Delta z_{0} + \Delta z_{DCR} + e,
	\label{eq:DCR_observe}
\end{equation}
where $ \Delta z_{(O-C)} $ is the positional  ($O-C$) (the observed minus computed) in ZD, and the computed position is made from some star catalog or some ephemeris. The color-unrelated effect of atmospheric refraction can be eliminated in $ \Delta z_{(O-C)} $ by fitting the plate model with a  $ 4^{th} $ order polynomial. In this way, the effects of distortion and color-unrelated atmospheric refraction can be well calibrated \citep{Lin2019, Guo2022RAA}. Specifically, we take the astrometric positions of some catalog for reference stars and perform differential astrometry. $ \Delta z_{(O-C)} $ usually changes during one night due to the change of both $ \Delta z_{0} $ and $ \Delta z_{DCR} $. For the first term in the right side of equation \ref{eq:DCR_observe}, $ \Delta z_{0} $ is the positional error in the direction of ZD. It is derived from the components of both  ($O-C$)s in right ascension (R.A.) and declination (Decl.) and projected into the direction of ZD. The ($O-C$)s in R.A. and Decl. are usually stable during a period of time. However, $ \Delta z_{0} $ of an object or a reference star usually changes during one night because of the change of its parallactic angle. $ \Delta z_{DCR} $ is the DCR effect, which is scaled as tangent of ZD. The light with shorter wavelength is more refracted, whereas longer wavelength less. $ e $ is the random error (e.g. the uncertainty of  measurement error). According to \citet{Lin2020MNRAS}, the DCR effect can be expressed as,
\begin{equation}
	\Delta z_{DCR} = \epsilon + \kappa \cdot CI \cdot tan(z),
	\label{eq:DCR_star}
\end{equation}
where $ \epsilon $ and $ \kappa $ are the DCR coefficients of a certain filter, $ CI $ is the color index (\textit{BP-RP} in Gaia system is used in this work) of the star and $ z $ is zenith distance. In equation \ref{eq:DCR_star}, since the absolute value of $\kappa$ is very small (less than 0.15 arcsec/mag for \textit{Null} filter), we can directly use the observed value of $ z $ for calculation. When we fit the plate model (e.g. $ 4^{th} $ order polynomial) and calculate $ \Delta z_{(O-C)} $, the DCR coefficient $ \epsilon $ will be a dummy coefficient because $ \epsilon $ (a constant) can be absorbed by the constant term of the polynomial. Hence, combining equations \ref{eq:DCR_observe} and \ref{eq:DCR_star}, the observed position in ZD can be expressed as,

\begin{equation}
	\Delta z_{(O-C)} = \Delta z_{0} +  \kappa \cdot CI \cdot tan(z) + e,
	\label{eq:updated_observe}
\end{equation}

\subsection{DCR Calibration from Reference Stars}
\label{section:2.3}
\par  We calibrate the DCR effect of the reference stars with a catalog-based method according to \citet{Lin2020MNRAS}. In the field of view, there should be at least 20 calibrated stars with a large range of known color index (e.g. 2.0 mag) in the field of view so that the DCR coefficient can be solved accurately. Specifically, for each frame, we first fit the plate model ($ 4^{th} $ order polynomial) using the matched stars between their pixel coordinates and standard coordinates in altitude-azimuth system based on the positional and photometric information of Gaia DR3 catalog \citep{2023AA_GaiaDR3}. Here, this polynomial contains the term $ \kappa \cdot CI \cdot tan(z) $ in the direction of ZD so that the DCR effect of stars can be eliminated. The positional errors of reference stars from the catalog ($ \Delta z_{0} $ in equation \ref{eq:updated_observe}) are considered negligible. In this way, the DCR coefficient $ \kappa$ for each frame can be solved. However, we can't directly obtain the catalog-matched color index of a solar system object from the catalog. In the following two subsections, we will elaborate on how to obtain the catalog-matched color index of such an object.

\begin{table*}
	\centering
	\small
	\caption{The observations of Himalia overview. The first column shows the observation date. The second column shows the filters used. The third and fourth columns show the number of frames acquired and exposure time corresponding to the filters used in the second column. The fifth column shows the range of ZD (zenith distance) during the observation. The telescopes used are shown in the sixth column. The seventh column shows the DCR coefficients $ \kappa $ with \textit{Cousins-I} filter. The last column shows the DCR coefficient $ \kappa' $ with \textit{Null} or \textit{Clear} filter. For the coefficient $ \kappa' $, the first four rows are the coefficients with \textit{Null} filter, while the last four with \textit{Clear} filter. }
	\begin{tabular}{cccccccr}
		\hline
		\hline
		\makebox[0.11\textwidth][c]{Date (UT)} &  \makebox[0.11\textwidth][c]{Filters} &  \makebox[0.09\textwidth][c]{Frames (No.)} & \makebox[0.09\textwidth][c]{ExpTime (s)} & \makebox[0.12\textwidth][c]{ZD (deg)} & \makebox[0.08\textwidth][c]{Telescope}  & \makebox[0.10\textwidth][c]{$\kappa$ (mas/mag)} &  \makebox[0.10\textwidth][c]{$\kappa'$ (mas/mag)} \\
		\hline
		2022-10-18 &\textit{Cousins-I / Null}  &  25 / 28    & 120 / 120      & 26.5 $\sim$ 51.0 &  \  \  \  1 m   & -5.8 $\pm$ 1.6  &  -95.8 $\pm$ 3.1 \\
		2022-10-19 &\textit{Cousins-I / Null}  &  37 / 36    & 120 / 120      & 27.6 $\sim$ 60.2 &  \  \  \  1 m     &  -9.4 $\pm$ 0.7 &  -100.7 $\pm$ 2.3 \\
		2022-10-20 &\textit{Cousins-I / Null}  &  27 / 27    & 120 / 120      & 26.9 $\sim$ 43.2 &  \  \  \  1 m     &  -5.7 $\pm$ 1.0  &  -118.9 $\pm$ 4.7 \\
		2022-10-21 &\textit{Cousins-I / Null}  &  18 / 17    & 120 / 120      & 26.9 $\sim$ 47.8 &  \  \  \  1 m     &  -8.6 $\pm$ 1.5  &  -110.4 $\pm$ 3.8 \\
		2022-11-22 &\textit{Cousins-I / Clear}& 54 / 55     & 120 /\ \ \ 80  & 29.3 $\sim$ 63.9 & 0.8 m &   -5.1 $\pm$ 0.5  &  -114.1 $\pm$ 1.2 \\
		2022-11-23 &\textit{Cousins-I / Clear}& 88 / 85     & 120 /\ \ \ 80  & 28.1 $\sim$ 64.7 & 0.8 m &   -5.3 $\pm$ 0.3  &  -117.3 $\pm$ 0.8 \\
		2022-11-29 &\textit{Cousins-I / Clear}& 89 / 91     & 120 /\ \ \ 80  & 28.1 $\sim$ 70.2 & 0.8 m  &   -5.0 $\pm$ 0.3  &  -105.8 $\pm$ 0.8 \\
		2022-11-30 &\textit{Cousins-I / Clear}& 90 / 90     & 120 /\ \ \ 80   & 28.1 $\sim$ 70.2 & 0.8 m &   -5.0 $\pm$ 0.3  &  -110.7 $\pm$ 0.9 \\
		\hline
	\end{tabular}
	\label{table:observations}
\end{table*}
\begin{figure*}
	\centering
	\includegraphics[width=1.00\textwidth]{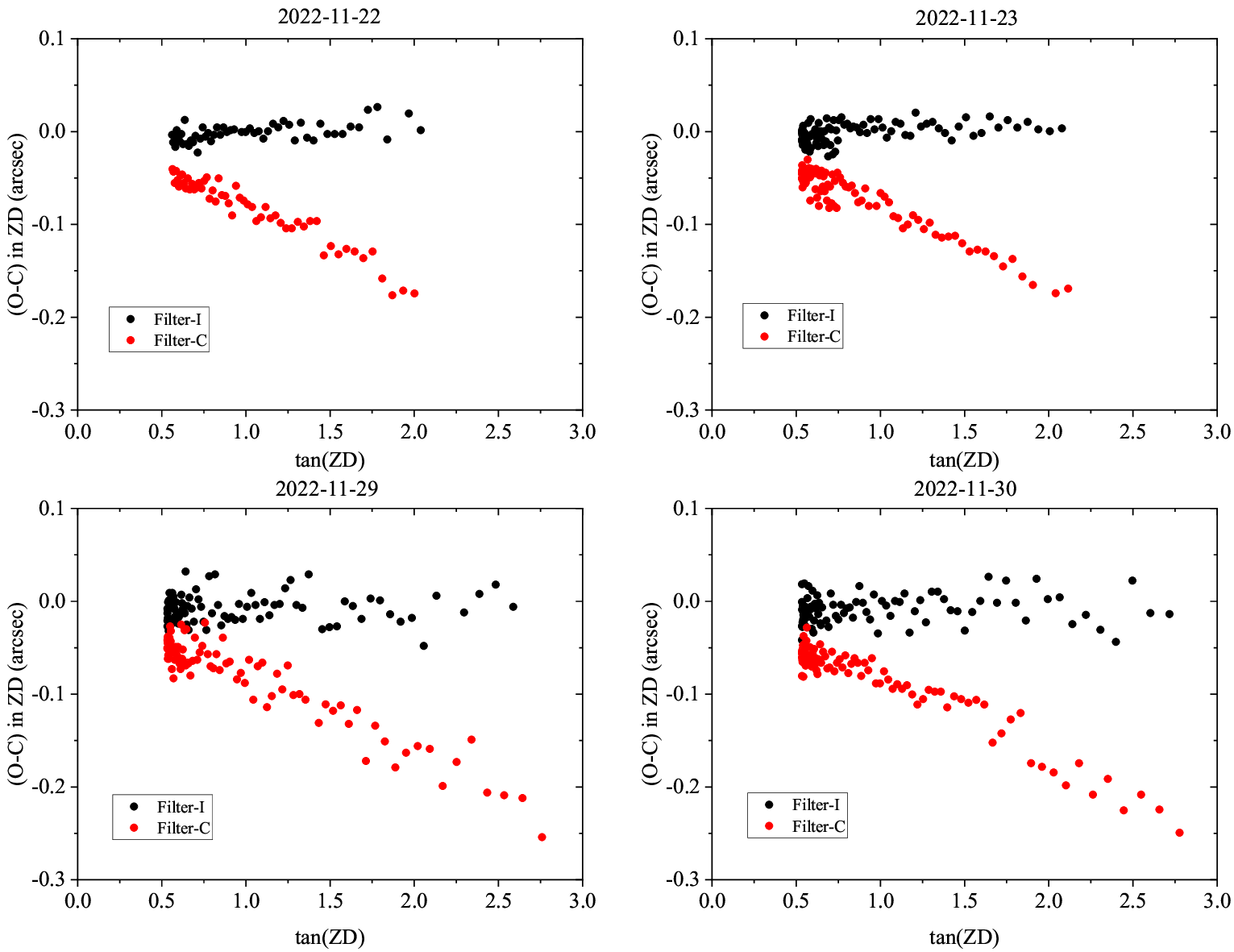}
	\caption{The four panels show the positional  ($O-C$)s of Himalia in ZD with tangent of ZD over 4 nights by the 0.8 m telescope. For each night, Himalia was observed with two different filters (\textit{Cousins-I} plotted in black and \textit{Clear} plotted in red) alternately with tangent of ZD.}
	\label{fig:tanzd}
\end{figure*}
\subsection{Measure the Color of an Object Using One Filter}
\label{section:2.4.1}
\par
\begin{figure*}
	\centering
	\includegraphics[width=1.00\textwidth]{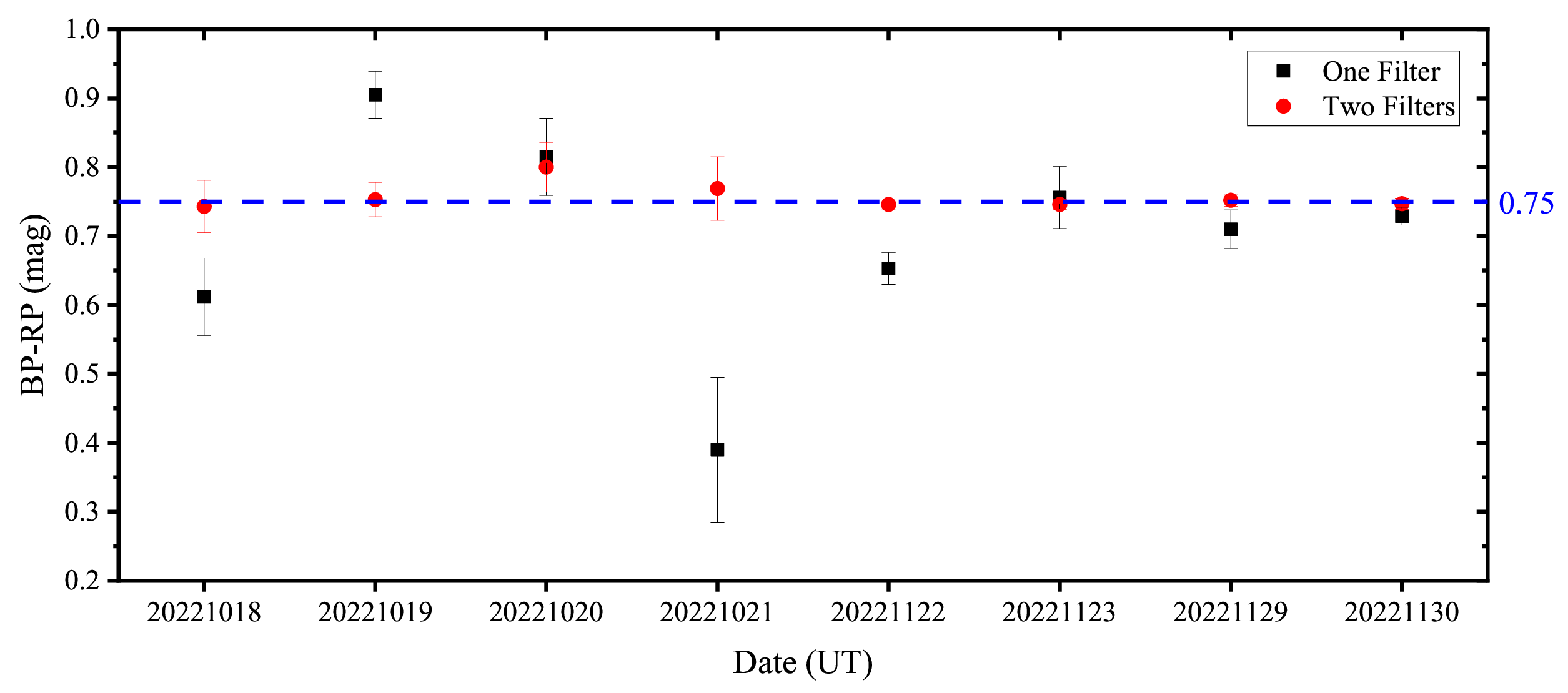}
	\caption{The figure shows the mean measured color indices $ BP-RP $ for each night using the observations taken by both the 1 m telescope and 0.8 m telescope. The red plots show the results measured by the method with two filters alternately, while the black plots show the results measured by the method with one filter. For clarity, a horizontal blue dash line ($ y = 0.75 $) is added in the figure.}
	\label{fig:two-filters}
\end{figure*}

 After DCR calibration from reference stars, we can conveniently derive the DCR coefficients $ \kappa $ of each filter. As for an object in solar system, in each frame, we can calculate $ \Delta z_{(O-C)} $ of the object. However, in equation \ref{eq:updated_observe}, neither $ \Delta z_{0} $ nor $ CI $ is known. $ \Delta z_{0} $ is the positional error which comes from the referenced ephemeris in R.A. and Decl.. For an object in solar system, $ \Delta z_{0} $ can not be negligible since the ephemeris may not provide the accurate position. According to the parallactic angle, the positional error in R.A. and Decl. can be projected into the direction of ZD.
Similar to equation \ref{eq:updated_observe}, for the positions of the object in altitude-azimuth system, we have the following expression (if we ignore the random error here),

\begin{equation}
	\left\{
	\begin{aligned}
		&\Delta z_{(O-C)}  =  \Delta z_{0} (q, \Delta \alpha \cos \delta, \Delta \delta)  + \kappa \cdot CI \cdot tan(z)\\
		&\Delta A_{(O-C)} \cos  h  = \Delta a_{0} (q, \Delta \alpha \cos \delta, \Delta \delta), \\
	\end{aligned}
	\right.
	\label{eq:one-filter}
\end{equation}
where $ q $ is the parallactic angle, $ \Delta \alpha \cos \delta  $ is the ($O-C$) in R.A., $ \Delta \delta $ is the ($O-C$) in Decl., and $ \Delta A_{(O-C)} \cos  h $ (also expressed as $ \Delta a_{0} $) is the ($O-C$) in azimuth. Both $ \Delta z_{0}$ and $  \Delta a_{0}  $ can be expressed as the function of $ q $, $ \Delta \alpha \cos \delta  $ and $ \Delta \delta $. The parallactic angle $ q $ can be calculated conveniently. In equation \ref{eq:one-filter}, $ \Delta \alpha \cos \delta  $, $ \Delta \delta $ and $ CI $ are all unknown and we can regard them as the constants during a period of time (e.g. during one night). Therefore, at least 3 observations are needed to solve the set of equations above. This method requires the observations with a large range of ZDs and large change of DCR effect for robust solution. It is obvious that this method it not convenient to measure the instant color index of the object, because we can only solve the mean DCR coefficient $ \kappa  $ and the mean color index. In practice, the uncertainty of the measured color index with this scheme is usually large (also see Figure \ref{fig:two-filters}) due to the large condition number. In addition, the solution might not be inaccurate due to the complex relationships of $ \Delta \alpha \cos \delta  $, $ \Delta \delta $ and $ CI $ (they might not be the constants during a short period of time). To accurately measure the instant color index, we can use another observation scheme with two filters alternately for measurement, which will be elaborated in the next subsection.

\subsection{Measure the Color of an Object Using Two Filters Alternately}
\label{section:2.4}
\par As shown in Subsection \ref{section:2.1}, we can derive the DCR coefficient $ \kappa $ of each filter. $ \Delta z_{(O-C)} $ of the object can be calculated. According to equation \ref{eq:updated_observe}, we can also solve $ CI $ (and $ \Delta z_{0} $) via at least two equations with different $ \Delta z_{(O-C)} $ and $ \kappa $ at the same moment. Namely, we can observe with at least two different filters to solve the color index $ CI $ (and $ \Delta z_{0} $). For a frame observed with one filter (denoted by \textit{filter-P}), we can calculate the $ \Delta z_{(O-C)} $ of the object, and as shown in equation \ref{eq:updated_observe}, $ \Delta z_{(O-C)}^P  = \Delta z_{0}^P + \kappa^P \cdot CI \cdot tan(z) $. We ignore the random errors during the derivation. Similarly, for a frame observed with another filter (denoted by \textit{filter-Q}), $ \Delta z_{(O-C)}^Q  = \Delta z_{0}^Q + \kappa^Q \cdot CI \cdot tan(z) $. If we alternately observe with the two filters, we can fit $ \Delta z_{(O-C)}^P $ or $ \Delta z_{(O-C)}^Q $ as the respective function with time. In this way, we can obtain $ \Delta z_{(O-C)}^P $ and $ \Delta z_{(O-C)}^Q $ at any moment during the period of observation. We do not need to fit such a function if the telescope can obtain two frames with two different filters simultaneously. Suppose at the moment of $ t $, the DCR effect with \textit{filter-P} and \textit{filter-Q} can be expressed as the following two equations.
\begin{equation}
	\Delta z_{(O-C)}^P(t) - \Delta z_{0}^P(t) =  \kappa^P \cdot CI \cdot tan(z(t)),
	\label{eq:DCR_A}
\end{equation}
\begin{equation}
	\Delta z_{(O-C)}^Q(t) - \Delta z_{0}^Q(t)  =   \kappa^Q \cdot CI \cdot tan(z(t)).
	\label{eq:DCR_B}
\end{equation}
In the two equations above, at the same moment of $ t $, $ \Delta z_{0}^P(t) $ equals $ \Delta z_{0}^Q(t) $. Let equation \ref{eq:DCR_B} minus equation \ref{eq:DCR_A}, for the same moment, we have the following expression,
\begin{equation}
	\Delta z_{(O-C)}^Q(t) - \Delta z_{(O-C)}^P(t) =   (\kappa^Q-\kappa^P) \cdot CI \cdot tan(z(t)).
	\label{eq:CI_minus}
\end{equation}
In this way, $ \Delta z_{0} $ in equation \ref{eq:updated_observe} can be eliminated at the moment of $ t $. For each observed frame with some filter (e.g. \textit{filter-P}), we can directly calculate the $ \Delta z_{(O-C)}^P(t)$ at the moment of $ t $ (the mid-time of the observed exposure frame). Then, the observed position with another filter (e.g. \textit{filter-Q}) $ \Delta z_{(O-C)}^Q(t)$ at the moment of $ t $ can be interpolated by the fitted function. According to equation \ref{eq:CI_minus}, all the parameters except $ CI $ can be obtained. Therefore, we can calculate the $ CI $ of the object at the moment of $ t $. The uncertainty of the measured color index at the moment of $ t $ can be approximately (ignore the uncertainty of $\kappa$) estimated as equation \ref{eq:error},
\begin{equation}
	\sigma^2_{CI} =  \dfrac{\sigma_P^2+\sigma_Q^2}{(\kappa^Q-\kappa^P)^2 \cdot \tan^2(z(t))},
	\label{eq:error}
\end{equation}
where $ \sigma_P $ and $ \sigma_Q $ are the positional error in ZD with \textit{filter-P} and \textit{filter-Q}, respectively. It can be seen that the uncertainty is mainly caused by the absolute value of difference of DCR coefficients $\lvert \kappa^Q-\kappa^P \rvert$ and the observed zenith distance. Hence, to obtain the accurate and precise color index, we should observe with large difference of DCR coefficients (e.g. using \textit{Cousins-I} and \textit{Null} filter alternately) and at large ZD. After we obtain the object's color indices and uncertainties, we can calculate a weighted (1/$ \sigma^2_{CI} $) average color index.

\par Figure \ref{fig:CI_errors} shows the DCR effect of an object change with ZD (left panel) and the estimated uncertainty changes with ZD (right panel). In the figure, we take the mean measured DCR coefficient and estimated uncertainties of color indices during the observation in this work as an example. The used observations in detail will be shown in Section \ref{section:3}. From the left panel, observed with \textit{Null} filter, the DCR effect of an object changes greatly with ZD (especially when the color index of the object is large). From the right panel, we can see that the uncertainty of color measurement is quite large at small ZD, but it is a good time to efficiently observe and perform astrometry with \textit{Null} filter. During the observation using the 1 m telescope (see the next section of observations), mainly due to the urban light pollution and poor weather conditions, the uncertainty of color measurement is larger than the 0.8 m telescope (also see the next section).

\section{Observations}
\label{section:3}
\par To verify the effectiveness of our method, we observe the jovian irregular satellite Himalia (the sixth satellite of Jupiter) for color measurement and astrometry over 8 nights by two different telescopes. The observations of Himalia overview are shown in Table \ref{table:observations}. Totally, 215 frames are obtained over 4 nights by the 1 m telescope at Kunming Station, Yunnan Observatory (IAU code: 286). And 642 frames are obtained over 4 nights by the 0.8 m telescope at Yaoan Station, Purple Mountain Observatory (IAU code: O49). During the observation by the 1 m telescope, the observations with consecutive time for one night may not be obtained due to the instrumental troubles and poor weather conditions. In addition, for the 1 m telescope, we could not observe the object with its ZD larger than 62$^\circ$ due to its dome occlusion. In contrast, observations with better quality are obtained by the 0.8 m telescope with the ZD as large to 70$^\circ$. For all the taken observations, at least 30 calibrated Gaia stars can be detected and matched in the field of view. As for the filters, we use \textit{Cousins-I} and \textit{Null} alternately by the 1 m telescope, while \textit{Cousins-I} and \textit{Clear} alternately by the 0.8 m telescope.

\section{Results}
\label{section:4}
\label{section:Image Processing}
After bias and flat-field corrections, for each frame, the two-dimensional Gaussian centering algorithm is used to measure the detected star images in pixel coordinate. For calibration, we match the detected stars with those in Gaia DR3 catalog \citep{2023AA_GaiaDR3} and obtain their astrometric and photometric information.

\subsection{Results of Color Index}
\label{section:Results of Color Index}
\par For the obtained observations of Himalia, according to Subsection \ref{section:2.3}, we solve the DCR coefficient $ \kappa $ of each filter, and the results are shown in Table \ref{table:observations}. For the same filter of the same telescope, empirically, the DCR coefficient $ \kappa $ is usually stable over several nights unless the poor weather conditions. To show the DCR effect of Himalia clearly, we calculate the ($O-C$) in ZD ($ \Delta z_{(O-C)} $ in equation (\ref{eq:updated_observe}) compared with JPL ephemeris\footnote{https://ssd.jpl.nasa.gov/horizons.cgi\#top} (\textit{JUP344} and \textit{DE441}). Figure \ref{fig:tanzd} shows the ($O-C$)s in ZD with the tangent of ZD as an example, alternately observed with \textit{Cousins-I} and \textit{Clear} filters by the 0.8 m telescope. For each panel, the  ($O-C$)s with \textit{Cousins-I} filter are almost near 0. Both the ($O-C$)s with \textit{Clear} and \textit{Cousins-I} filter are almost linear with the tangent of ZD. If we fit their respective linear function, they will approximately intersect at zenith ($ tan(ZD)=0 $). This numeral relationship is also shown in equation \ref{eq:updated_observe}.

\begin{table}
	\centering
	\small
	\caption{The table shows the color indices $ BP-RP $ and its uncertainty of Himalia. As for the work from other researchers, we have transformed the original measured color indices to the $ BP-RP $ system according to Gaia Data Release 3 Documentation. The color indices in this work are measured by the observations in Table \ref{table:observations}.}
	\begin{tabular}{cccc}
		\hline
		\hline
		\makebox[0.015\textwidth][c]{Work} &  \makebox[0.10\textwidth][c]{Observations} & \makebox[0.09\textwidth][c]{BP-RP} & \makebox[0.09\textwidth][c]{Uncertainty}   \\
		&   & \makebox[0.09\textwidth][c]{(mag)} & \makebox[0.09\textwidth][c]{(mag)}   \\
		\hline
		\multirow{3}{*}{Others} & \citet{Degewij1980Icar} & 0.78 $ \  $    & 0.08  $ \  $  \\
		&  \citet{Luu1991AJ}              & 0.75 $ \  $   & 0.07  $ \  $  \\
		&  \citet{Rettig2001Icar}  & 0.75 $ \  $    & 0.09  $ \  $  \\
		\hline
		\multirow{8}{*}{This work}   & 2022-10-18   & 0.743   & 0.038  \\
		& 2022-10-19   & 0.753   & 0.025  \\
		& 2022-10-20   & 0.800   & 0.036   \\
		& 2022-10-21   & 0.769   & 0.046   \\
		& 2022-11-22   & 0.746   & 0.008   \\		
		& 2022-11-23   & 0.746   & 0.007   \\	
		& 2022-11-29   & 0.752   & 0.009   \\	
		& 2022-11-30   & 0.747   & 0.008   \\	
		\hline
		\multirow{3}{*}{Total}         &   Subtotal (1 m)  & 0.763   & 0.017   \\	
		& Subtotal (0.8 m) & 0.748   & 0.004   \\	
		& Total (all)     & 0.750   & 0.004   \\	
		\hline
	\end{tabular}
	\label{table:total_color}
\end{table}

\par We calculate the color indices \textit{BP-RP} (in Gaia photometric system) of Himalia for the observations of each night using two methods (using one filter and two filters alternately). As for the method of using one filter, only the observations with \textit{Null} or \textit{Clear} filter are used for calculation. We obtain a color measurement result for the observations over one night. Observations with \textit{Cousins-I} filter will have influence on the accuracy of the results because of the large condition number. As for the method of using two filters, we use all the taken observations as shown in Table \ref{table:observations}. The comparison of the results with two methods are shown in Figure \ref{fig:two-filters}. Obviously, we can derive more accurate and precise results using the method of two filters alternately.

\par
\begin{figure*}
	\centering
	\includegraphics[width=0.90\textwidth]{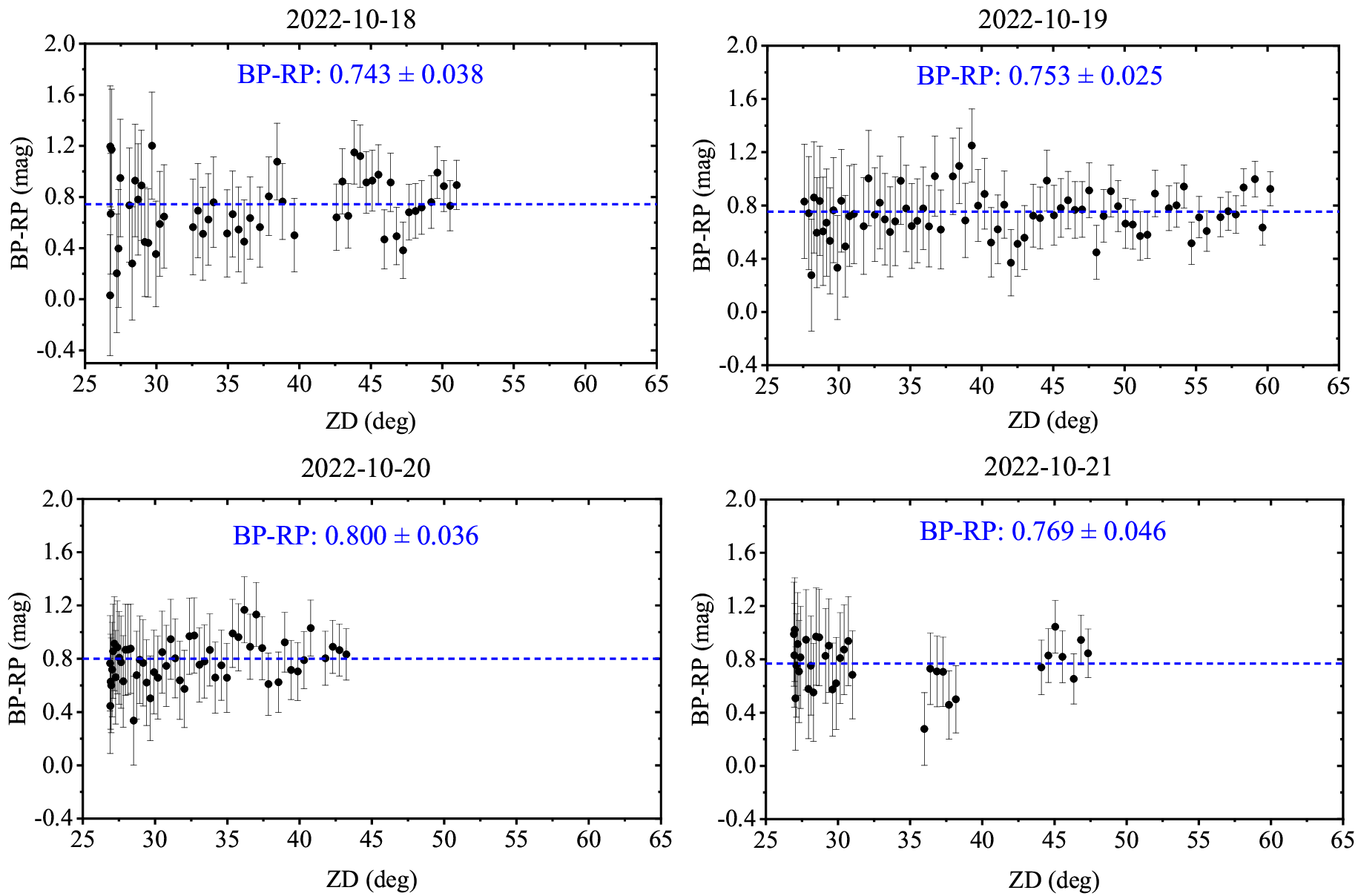}
	\caption{The four panels show the measured color indices $ BP-RP $ for each night with ZD using the observations taken by the 1 m telescope. The weighted mean color indices are shown in blue dash lines. The values of the weighted mean color indices and their uncertainties are shown at the top of each panel.}
	\label{fig:Color_KM}
\end{figure*}

\begin{figure*}
	\centering
	\includegraphics[width=0.90\textwidth]{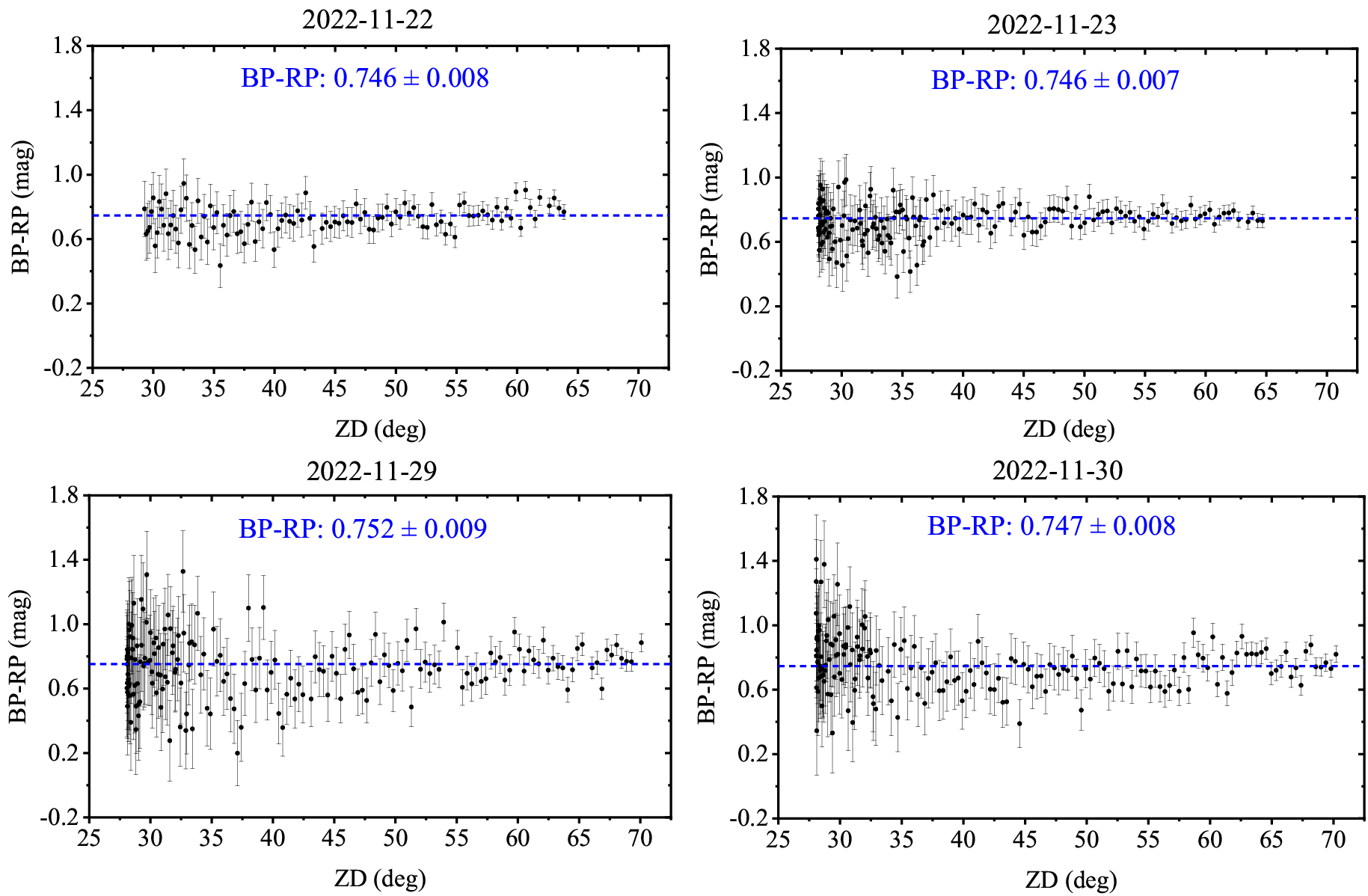}
	\caption{The four panels show the measured color indices $ BP-RP $ for each night with ZD using the observations taken by the 0.8 m telescope. The weighted mean color indices are shown in blue dash lines. The values of the weighted mean color indices and their uncertainties are shown at the top of each panel.}
	\label{fig:Color_YA}
\end{figure*}

 Figures \ref{fig:Color_KM} and \ref{fig:Color_YA} show the color measurement results of Himalia using the method of two filters alternately by the 1 m and 0.8 m telescope, respectively. From Figure \ref{fig:Color_KM}, especially for the bottom-right panel, due to the bad weather, only 7 frames (for either \textit{Cousins-I} or \textit{Null} filter) with each ZD greater than 40$^\circ$ are taken. In contrast, for the four panels in Figure \ref{fig:Color_YA} observed by the 0.8 m telescope, the observations are consecutive for each night, and the observations at large ZD are also obtained. On Nov 22 and 23, 2022, observations are obtained at the large ZD of $\sim$ 65$^\circ$. While on Nov 29 and 30, 2022, the ZD is even as larger as $\sim$ 70$^\circ$. Therefore, the uncertainties of the mean color indices by the 0.8 m telescope are smaller.

\par To verify the measured color index in this work, we compared it with other references. For comparison, we have changed their original measured color indices into the \textit{BP-RP} system according to Gaia Data Release 3 Documentation\footnote{\it https://gea.esac.esa.int/archive/documentation/GDR3/}. The comparison results are shown in Table \ref{table:total_color}. The color indices we measured in this work using the method of two filters alternately show the consistency for each night, and also with other works. Totally, the mean color index \textit{BP-RP} of Himalia measured in this work is 0.750 $\pm$ 0.004 magnitude. The reasons why we spend 8 nights using two different telescopes to measure the color index are shown in the following three aspects. a) to verify whether we can obtain the same result in different nights; b) to verify whether we can obtain the same result using different telescopes; c) to verify the uncertainty of the measured color index with respect to zenith distance. If we want to measure the color index in a short time, we can just observe the target at large zenith distances.

\begin{table*}
	\centering
	\scriptsize
	\caption{The table shows the $ \langle O-C \rangle $s (mean the observed minus computed) and the standard deviations (SD) of Himalia before and after DCR calibration with different filters. The first column shows the observation date. The second and third columns show the results with \textit{Cousins-I} filter before DCR calibration in right ascension and declination, respectively. The fourth and  fifth columns show the results with \textit{Null} or \textit{Clear} filter before DCR calibration. The last four columns show the reduced results after DCR calibration with different filters.}
	\begin{tabular}{rrrrrrrrr}
		\hline
		\hline
		\makebox[0.06\textwidth][c]{Date (UT)} &  \makebox[0.09\textwidth][c]{R.A. (I)} & \makebox[0.09\textwidth][c]{Decl. (I)} & \makebox[0.09\textwidth][c]{R.A. (N/C)} & \makebox[0.09\textwidth][c]{Decl. (N/C)} & \makebox[0.09\textwidth][c]{R.A. (I)} & \makebox[0.09\textwidth][c]{Decl. (I)} & \makebox[0.09\textwidth][c]{R.A. (N/C)}  & \makebox[0.09\textwidth][c]{Decl. (N/C)} \\
		& \multicolumn{4}{c}{$ \langle O-C \rangle  $ $\pm$ SD (before DCR calibration) in arcsec} & \multicolumn{4}{c}{$ \langle O-C \rangle  $ $\pm$ SD (after DCR calibration) in arcsec} \\
		\hline
		2022-10-18 & 0.019 $\pm$ 0.019 &  0.006 $\pm$ 0.012    & -0.003 $\pm $ 0.030   & 0.049 $\pm$ 0.012 &  0.020 $\pm$ 0.019    & 0.005 $\pm$ 0.012   & 0.029 $\pm$ 0.018 & 0.010 $\pm$ 0.013 \\
		2022-10-19 & 0.017 $\pm$ 0.015 &  -0.004 $\pm$ 0.014    & 0.010 $\pm $ 0.015   & -0.001 $\pm$ 0.018 &  0.015 $\pm$ 0.015    & -0.002 $\pm$ 0.015   & 0.019 $\pm$ 0.015 & -0.008 $\pm$ 0.018 \\
		2022-10-20 & 0.012 $\pm$ 0.016 &  -0.004 $\pm$ 0.010    & 0.003 $\pm $ 0.013   & 0.015 $\pm$ 0.012 &  0.012 $\pm$ 0.016    & -0.005 $\pm$ 0.010   & 0.017 $\pm$ 0.014 & -0.007 $\pm$ 0.013 \\
		2022-10-21 & 0.024 $\pm$ 0.019 &  0.008 $\pm$ 0.012    & 0.023 $\pm$ 0.020   & 0.030 $\pm$ 0.025 &  0.024 $\pm$ 0.019    & 0.008 $\pm$ 0.012   & 0.034 $\pm$ 0.018 & 0.012 $\pm$ 0.026 \\		
		2022-11-22 & 0.019 $\pm$ 0.008 &  -0.008 $\pm$ 0.009    & 0.090 $\pm$ 0.041   & 0.059 $\pm$ 0.015 &  0.017 $\pm$ 0.007    & -0.009 $\pm$ 0.009   & 0.014 $\pm$ 0.007 & -0.007 $\pm$ 0.009 \\				
		2022-11-23 & 0.020 $\pm$ 0.006 &  -0.001 $\pm$ 0.007    & 0.063 $\pm$ 0.058   & 0.058 $\pm$ 0.014 &  0.018 $\pm$ 0.006    & -0.004 $\pm$ 0.007   & 0.018 $\pm$ 0.005 & -0.003 $\pm$ 0.008 \\		
		2022-11-29 & 0.014 $\pm$ 0.013 &  -0.009 $\pm$ 0.020    & 0.055 $\pm$ 0.047   & 0.031 $\pm$ 0.020 &  0.012 $\pm$ 0.013    & -0.010 $\pm$ 0.020   & 0.015 $\pm$ 0.013 & -0.013 $\pm$ 0.022 \\		
		2022-11-30 & 0.014 $\pm$ 0.014 &  -0.011 $\pm$ 0.018    & 0.061 $\pm$ 0.054   & 0.043 $\pm$ 0.019 &  0.012 $\pm$ 0.013    & -0.013 $\pm$ 0.018   & 0.011 $\pm$ 0.012 & -0.011 $\pm$ 0.013 \\				
		\hline
	\end{tabular}
	\label{table:results_rade}
\end{table*}

\par
\begin{figure*}
	\centering
	\includegraphics[width=1.00\textwidth]{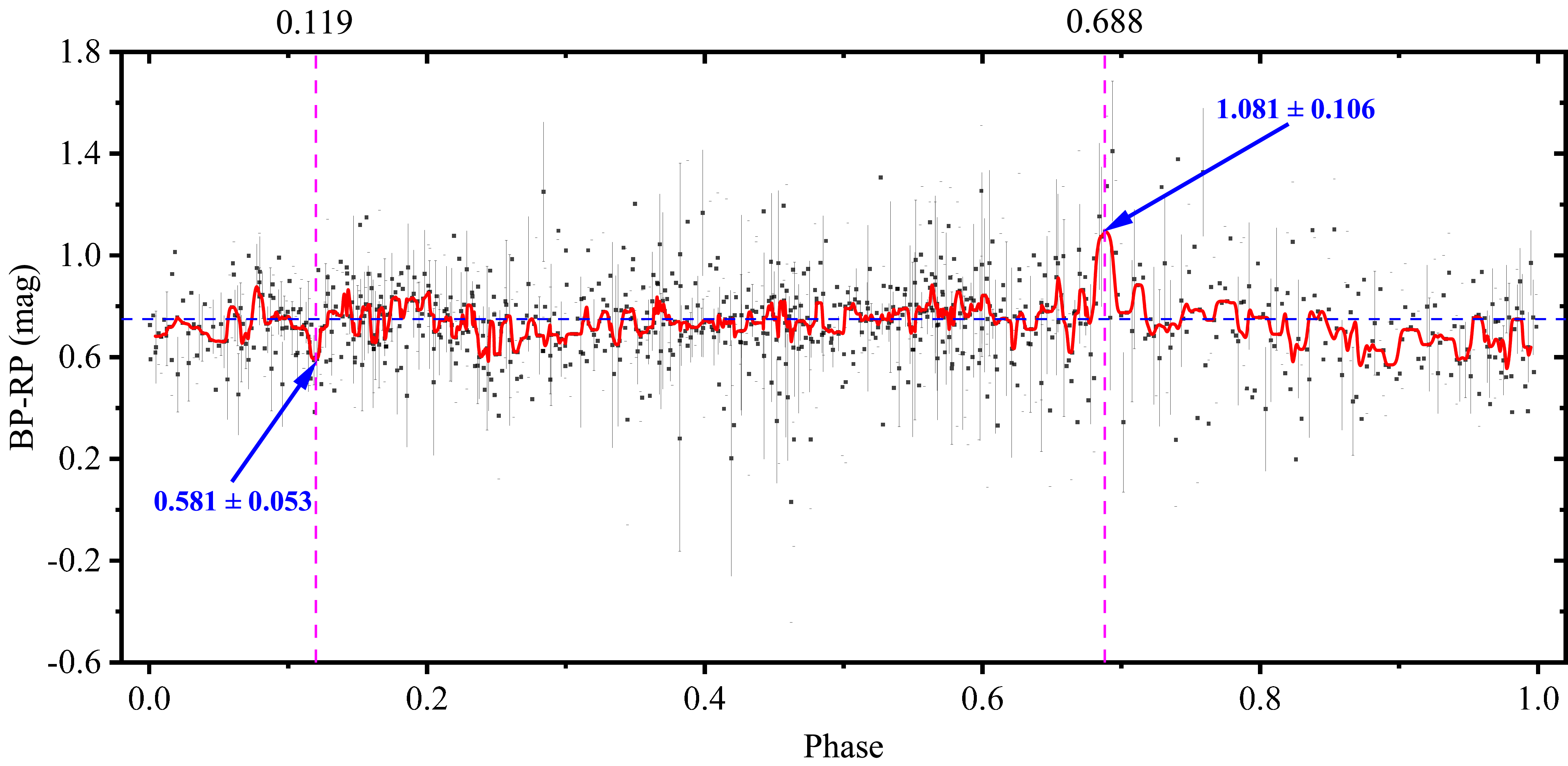}
	\caption{This figure shows the phased color indices of Himalia using all the measured results in this work. The phase is calculated according to its rotation period 7.7819h \citep{Pilcher2012Icar}, and the light time is also taken into account. The blue dash line is the weighted mean color index 0.750. The red line is the smoothed color index using an average filter for the nearest seven plots. The two pink vertical lines show the two obvious places with great change of the phased color indices, and the corresponding values attached with blue arrows are the smoothed color index $\pm$ 1$\sigma$. }
	\label{fig:color_phase}
\end{figure*}

 At present, the change of Himalia's color within a short time (e.g. the change with its rotation) has not been explored. The color indices of Himalia measured by other works are regarded as its instant color, which are only measured by several frames on one night. Its color index is usually considered as a constant. According to all the color measurement results in this work, we have them rotational phased based on its rotation period 7.7819h \citep{Pilcher2012Icar}, which is shown in Figure \ref{fig:color_phase}. We used an average filter for the nearest seven plots to smooth the color index with its rotational phase. According to the smoothed phased color index, two places are found with their color indices exceeding the mean $\pm$ 3$\sigma$. One is near the phase of 0.119 where the color index is small to 0.581 $\pm$ 0.053 magnitude. The other is near the phase of 0.688 and the color index is up to 1.081 $\pm$ 0.106 magnitude. The change of color index with its rotational phase might indicate the uneven material distributions on the surface.

\subsection{Results of Astrometry in Equatorial Coordinate}
\label{section:Results of Astrometry}
To calibrate the DCR effect of both reference stars and Himalia in data reduction, according to the astrometric information from the star catalog, we firstly calculate the standard coordinate of each catalog-matched star at the observational epoch through the central projection \citep{Green1985}. Here, the astrometric positions of stars are taken in equatorial coordinate. Then, the least squares scheme is used to solve the plate model with a weighted 4$ ^{th} $ order polynomial \citep{Lin2019}, which also contains the terms of DCR coefficient, color index, parallactic angle and ZD. According to the solved DCR coefficient $ \kappa $ and color indices (the color indices of reference stars are from Gaia DR3 catalog and the mean color index of Himalia is solved in this work), the observed positions of both reference stars and the Himalia can be calculated via the solved plate model. The calculate position of Himalia is computed by the latest version of JPL ephemeris and we can calculate the positional ($ O-C $) of Himalia.

\par  The results before and after the DCR calibration are shown in Table \ref{table:results_rade}. Here, the astrometric results before DCR calibration are calculated using a weighted 4$ ^{th} $ order polynomial without containing the items of DCR calibration to fit the plate model. From our results, there are only a few differences in the observations with \textit{Cousins-I} filter after DCR calibration, which also shows that using \textit{Cousins-I} filter can effectively reduce the DCR effect. As for the DCR effect of \textit{Null} and \textit{Clear} filters, it depends on the difference color indices between the reference stars and the object as well as the position distributions of reference stars in the field of view. For the results after DCR calibration, we can derive similar astrometric results for both observed with \textit{Cousins-I}, \textit{Null} or \textit{Clear} filters. The reduction results after DCR calibration show that the standard deviations of $ (O-C)$s (the observed minus computed) compared with JPL ephemeris is estimated at 0.005 $\sim$ 0.019 and 0.007 $\sim$ 0.026 arcsec in R.A. and Decl., respectively.

\section{Discussion}
\label{section:5}

\par Compared with the two methods (using one filter and two filters alternately), we can obtain more accurate and precise color index by the method of two filters alternately. In addition, we can conveniently obtain the instant color index of an object using two filters alternately. However, we should concentrate on the telescope focus. A proper focusing should be found for both filters. Namely, a good focus for \textit{Cousins-I} filter might not fit \textit{Null} or \textit{Clear} filter. In contrast, we do not need to worry about the focus using only a filter.  As for the method of using two filters alternately, there is a contradiction between the color and position measurement. When observed at larger ZD, although we can derive more accurate color index of an object with the same positional precision, the measurement error of the pixel position would be larger mainly due to the greater atmospheric extinction. From the observations used in this work, we find it works well with the ZD up to $\sim$ 70$^\circ$, and the effectiveness of observing with larger ZD should be further explored. The purpose of our work is to effectively calibrate the DCR effect in astrometry. Therefore, we focus less on what kind of color indices are used. To measure the color of an object, this method requires the corresponding color indices (e.g. \textit{B-V}) of enough reference stars in the field of view. However, we cannot find the corresponding color indices (e.g. \textit{B-V}) of most reference stars. In contrast, \textit{Gaia} catalog provides a large number of reference stars with accurate color indices, which is more convenient to our work. Therefore, this method can effectively measure the color indices in Gaia system. In this work, we measure the color index \textit{BP-RP} with larger numerical range of reference stars compared with \textit{G-RP} and \textit{BP-G} so that the solution of DCR coefficient is more accurate and precise.

To measure the color index precisely using two filters alternately, the observations should be made according to the requirements as follows. Firstly, the two filters used should cause large difference of DCR effect. For example, we can alternately observe with \textit{Cousins-I} filter (the small DCR effect) and \textit{Null} filter (the large DCR effect). Secondly, we should observe at large ZD (90$^\circ$ - altitude), e.g. larger than 40$^\circ$, no matter when located before or after the meridian. Thirdly, enough frames should be obtained to reduce the influence of random errors, and we suggest at least 15 frames for each filter. Fourthly, there should be more than 20 reference stars with a large range of color index (e.g. 2.0 mag) in the field of view to solve the DCR coefficients accurately.

\par DCR has been regarded as a negative effect for ground-based observation especially for astrometry, since it makes the accurate positional measurement difficult. However, in this work, we take advantage of the DCR effect and measure the  color of an object, which also builds a link between astrometry and photometry. In fact, two kinds of differential astrometry are performed based on the catalog. One is the plate model between the pixel and equatorial standard coordinate, which shows the differential positions of stars in the field of view. Another is the differential relationship of DCR effect between the reference stars and the target (solar system object). Namely, the Earth's atmosphere will have the same positional influence on both the reference stars and the target. Although there are some effects caused by the atmosphere (e.g. diffused effect), we can obtain the accurate color index of the target via differential measurement. In addition, the atmospheric conditions (e.g. the atmospheric transparency and thin clouds) might decrease the SNRs of stars. However, it will not influence the positional accuracy, but will influence the positional uncertainty. For color index measurement, the positional uncertainty has been taken into account in equation (\ref{eq:error}).

\par In order to explore whether this method can accurately measure the objects with different color indices. For testing, we can assume that the color indices of some reference stars (with different colors) are unknown. Then, we measure their color indices and make a comparison with those in \textit{Gaia} catalog. Specifically, these reference stars for testing are not used for DCR calibration, which are regarded as the objects with unknown color indices. Table \ref{table:color indices of stars} shows the testing results using the observations with \textit{Clear} filter on Nov, 23, 2022. Here, we assume that the positional errors in \textit{Gaia} catalog are negligible. The results show that our method also works when measuring the objects with other color indices.

\begin{table}
	\centering
	\footnotesize
	\caption{The table shows measured color indices $ BP-RP $ of the reference stars in comparison with those in \textit{Gaia} catalog. The observations on Nov 23, 2022 are used for testing. The listed stars are not used for DCR calibration. The first column shows the \textit{Gaia} source ID of stars. The second column shows the G-magnitude of stars. The third column lists the results of measured color indices in this work, together with the uncertainty. The last column shows the color indices and the uncertainty provided by \textit{Gaia} catalog.}
	\begin{tabular}{cccc}
		\hline
		\hline
		\makebox[0.10\textwidth][c]{Source ID} &  \makebox[0.06\textwidth][c]{Magnitude} & \makebox[0.09\textwidth][c]{Measured CI} & \makebox[0.09\textwidth][c]{\textit{Gaia} CI}   \\
		&   & \makebox[0.05\textwidth][c]{(mag)} & \makebox[0.09\textwidth][c]{(mag)}   \\
		\hline
		2449158753553218816 & 14.047   & 1.198 $\pm$ 0.016   & 1.185 $\pm$ 0.007  \\
		2449162704923136512 & 14.656   & 2.157 $\pm$ 0.017   & 2.160 $\pm$ 0.007   \\
		2449162670563388544 & 14.992   & 0.855 $\pm$ 0.019   & 0.826 $\pm$ 0.007   \\
		2449157791480546048	& 15.116   & 0.996 $\pm$ 0.016   & 0.983 $\pm$ 0.007   \\		
		2449151950325022848 & 15.447   & 1.300 $\pm$ 0.018   & 1.296 $\pm$ 0.008   \\	
		2449151778526331904	& 15.541   & 1.032 $\pm$ 0.018   & 1.024 $\pm$ 0.008   \\	
		2449151709806855424	& 15.771   & 0.952 $\pm$ 0.018   & 0.989 $\pm$ 0.008   \\		
		2448963865117864064	& 15.789   & 1.187 $\pm$ 0.018   & 1.197 $\pm$ 0.008   \\	
		2449151434928925312	& 16.397   & 0.838 $\pm$ 0.020   & 0.864 $\pm$ 0.010   \\		
		2449159784345367680	& 16.608   & 0.854 $\pm$ 0.021   & 0.850 $\pm$ 0.012   \\
		\hline
	\end{tabular}
	\label{table:color indices of stars}
\end{table}

\par Compared with obtaining color index by absolute photometry, our method (using two filters alternately) may be inefficient when observed at small ZD. There are two aspects to elaborate the advantages of our method. One is that our method is not so demanding for weather conditions or well-calibrated instruments. For example, absolute photometry might not work well if the thin or curve clouds exist in some sky areas or flat fields are not taken. In some sense, the photometric nights are rare. In contrast, the cases above have smaller influence when using our method. Another is that the catalog-matched color index can be directly obtained via our method, and it serves for DCR calibration in astrometry. For example, if we would like to calibrate the DCR effect of Himalia based on the Gaia star catalog, the color indices of both stars and Himalia in the Gaia photometric system should be obtained. We can obtain the precise color indices \textit{BP-RP} of stars from Gaia catalog. As for Himalia, we can directly derive the color index of Himalia in \textit{BP-RP} system using this method. However, the measured color index system by absolute photometry is based on the photometric system of the filters attached. The transformation to the catalog-matched color index system may cause a large error (e.g. 0.02$\sim$0.20 mag), which will cause the positional errors of up to tens of milli-arcseconds with \textit{Null} filter after DCR calibration.

\section{Conclusion}
\label{section:6}
\par In this work, we present an astrometric method to measure the catalog-matched color index of an object via the DCR effect. Once the color index has been derived, we can observe it with \textit{Null} or \textit{Clear} filter during a large range of ZDs and perform DCR calibration, by which the limiting magnitude and the efficiency of the telescope can be significantly improved. In addition, this method builds a link between astrometry and photometry. Take the jovian satellite Himalia as an example, we measure its color indices and positions using the observations over 8 nights by two telescopes. To verify whether this method can measure the objects with different color indices, the stars with different color indices are measured for testing. Totally our results show that the mean color index $ BP-RP $ of Himalia is 0.750 $\pm$ 0.004 magnitude. After DCR calibration, the standard deviations of the positional $ (O-C)$s (the observed minus computed) compared with JPL ephemeris are estimated at 0.005 $\sim$ 0.019 and 0.007 $\sim$ 0.026 arcsec in R.A. and Decl. for each night, respectively.

\section*{Acknowledgements}

This work was supported by the National Key R\&D Program of China (Grant No. 2022YFE0116800), by the China Manned Space Project (Grant No. CMS-CSST-2021-B08), by the National Natural Science Foundation of China (Grant Nos. 11873026, 11273014) and Joint Research Fund in Astronomy (Grant No. U1431227). We thank the anonymous reviewer who provided us with valuable comments, Dr. Lin F. R. who had a helpful discussion with us and Mr. Cao J. L. who helped us match the observations. This work has made use of data from the European Space Agency (ESA) mission Gaia (\url {https://www. cosmos.esa.int/gaia}), processed by the Gaia Data Processing and Analysis Consortium (DPAC; \url{hppts://www.cosmos.esa.int /web/gaia/dpac/consortium}). Funding for the DPAC has been provided by national institutions, in particular the institutions participating in the Gaia Multilateral Agreement.

\section*{Data Availability}

The data underlying this article will be shared on reasonable request to the corresponding author.



\bibliographystyle{mnras}
\bibliography{example} 




\appendix


\bsp	
\label{lastpage}
\end{document}